\documentclass[sigconf, 10pt]{acmart}

\usepackage[english]{babel}
\usepackage{blindtext}

\renewcommand\footnotetextcopyrightpermission[1]{} 

\begin{document}

\title{Towards End-to-End Quality-of-Service by Domain-Level Routing and Forwarding}


\author{Haoyu Song}
 \affiliation{%
   \institution{Futurewei Technologies}
   \city{Santa Clara} 
   \state{USA} 
 }
 \email{haoyu.song@futurewei.com}

\renewcommand{\shortauthors}{H. Song}

\begin{abstract}

We propose to raise network domain to the first-class citizen in packet routing and forwarding. By making domain explicit in packet header, it can be held responsible for committed services and quality of service. Since the service is easier to be monitored, verified, and compensated, a new business model catering for emerging cross-domain, high-value, and performance-sensitive applications can be fostered. The domain-level routing and forwarding essentially create a L3.5 in the network protocol stack. We extend IPv6 with new routing headers to support two types of forwarding approaches. One is based on the next domain lookup and the other realizes the domain-level source routing. We describe various design details covering header format, forwarding behavior, routing protocol, DNS, and OAM. Several use cases are presented to demonstrate the scheme potential. The design is compatible with today's Internet architecture and incrementally deployable on existing infrastructure with only moderate updates needed. We expect the preliminary work can trigger more discussion and detailed design in the research and Internet standard communities.    

\end{abstract}

\pagestyle{plain}
\maketitle

\section{Introduction}

Emerging value-added network functions (e.g., INT~\cite{int2021} and SFC \cite{sfc2016}) are often claimed to be used in only a limited domain to secure their deployability. This is not because that they are technically incapable of being used in a wide area network crossing multiple domains, but because there is a lack of incentives for cross-domain interoperability.  Unfortunately, without the end-to-end usage, the value of such functions is also limited. This is not a new issue. It has long been recognized that inter-domain quality of service (QoS) is difficult to achieve due to economic and security reasons~\cite{rfc-iab}, and such efforts have even been largely abandoned. However, today, driven by emerging applications~\cite{itu2019net2030} (e.g., metaverse, live 3D-video for sports and entertainment events, and remote interactive operations, which usually require the global reach), the need for inter-domain QoS is becoming more urgent than before. The requirements of these applications are not limited to just high bandwidth and low latency in a statistical sense. They often have deterministic performance target with strict QoS criteria (e.g., zero packet drop guarantee, a constant bit rate, and a fixed latency budget). The Internet service providers must collaborate to achieve the QoS and service goals. If any involved one does not cooperate and behave, it makes no sense for the others to do anything better than the best-effort service as being practiced today, which essentially means no other guarantee beyond connectivity. Therefore, we believe the support of inter-domain QoS and cross-domain network functions is so important that it deserves further investigation. The dilemma of inter-domain cooperation must be cracked.

We need to provide right incentives for network service providers to foster a viable business model that allows multiple service providers to work in concert with each other to provide certain network-wide value-added services or QoS guarantee. In this paper we do not intend to discuss the business aspects of this issue but study the technical means to facilitate the effort to establish a business model which can be adopted to benefit both end users and service providers. The question is why the service providers are not motivated to provide inter-domain services and QoS before? The answer could be multi-faceted. But an important reason is that the service providers or domains are invisible from end host’s perspective and thus they cannot be held accountable even if they will or are able to do better.

While the routing architecture in today’s Internet is hierarchical (i.e., with the differentiation of Exterior Gateway Protocol (EGP) and Interior Gateway Protocol (IGP)), the resulting forwarding table is flat. The Autonomous Systems (AS), which play an important role in the Internet routing, are no longer visible after the routing table is computed. At any node in the network, a next hop is all needed for forwarding a packet towards its destination. Since the domains and domain boundaries are invisible from the end-host’s perspective, and the domain information is ignored by forwarding, it is difficult to measure and audit per-domain behavior and performance, and difficult to attribute any issue to any service provider. Simply put, if users have no visibility to the domains their traffic goes through, they cannot exert any effective control on it; similarly, if service providers have no clue whether the other domains behave, they are not motivated to react to any control.

Source routing offers users the capability to select the actual forwarding paths for their traffic. SRv6~\cite{rfc-srv6} has become a main-stream source-routing technology which works on the IPv6 data plane. However, without a clear delineation of responsibility, service providers would feel reluctant to support End-to-End (E2E) source routing because the designated routes may break the established peering relationship and yet provide no tangible business benefits.  

The solution is to raise the status of service provider (or domain) to the first-class citizen in packet forwarding.   It has been observed that the different stakeholders on the Internet have competing interests~\cite{tussle2002sigcomm}. The tussles exist between users and service providers and among service providers. It suggests that “the Internet should support a mechanism for choice of source routing that would permit a customer to control the path of his packets \emph{at the level of providers}”. 

It becomes clear that the providers must be visible to end users at the packet level. But how?  The answer is to introduce L3.5 in the network protocol layer stack and make domain the L3.5 entity. The vision of L3.5 has been articulated in~\cite{trotsky2019sigcomm}. It insightfully points out there is an unfortunate missing layer in the network protocol stack due to the early evolution of the Internet. The current protocol layer stack is formed before the advent of the Autonomous Systems. As the Internet scales out, multiple service providers and multiple roles (e.g., core, transit, and stub) are needed, so AS came into being, creating a new layer above IP effectively but which was largely ignored (except for routing).   

The authors in~\cite{trotsky2019sigcomm} however consider the introduction of L3.5 as a breakthrough to free the evolution of L3 because a new overlay can hide the underlay details and hence multiple different types of underlay (e.g., L3 protocols) can be supported. While we appreciate the insight, we hold a more modest opinion toward the actual usage of the new L3.5. It is not clear to us whether or not any new L3 protocol other than IP is ever needed, but nevertheless the introduction of L3.5 can bring the domain visibility to the forwarding plane. We believe this is a stronger motivation that makes the idea of L3.5 more compelling. Therefore, we focus on empowering IP-based Internet and try to extend it with a new domain layer to help service providers collaborate and provide better support of E2E service and QoS. 
The domain layer allows the network performance and service level to be monitored at the service provider level. Because of this, contributor and culprit can be clearly identified, and service providers are easier to be compensated and therefore are more likely to collaborate for some common goals.


\begin{figure}[tb]
\centering
\includegraphics[width=0.13\textwidth]{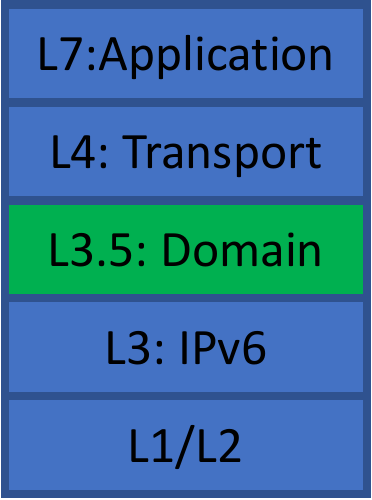}
\caption{Protocol stack with a new L3.5 layer.}
\label{fig:stack}
\end{figure}

This paper provides a design for IPv6 domain level routing and forwarding based on an abstract network model in which each domain is virtualized as a ``router". 
The remainder of the paper is organized as follows: Section~2 presents the domain-level routing architecture with the abstract network model, Section~3 details the system design which covers various technical aspects enabling the domain-level routing and forwarding, Section 4 provides several use cases to demonstrate the potential of the new design, Section~5 summarizes the relate work, Section 6 analyzes the system cost and discusses the future work, and finally, Section 7 concludes the work.       

\section{Architecture}

IPv6 is set to dominate the Internet, replacing IPv4 as the new narrow waist. According to Google statistics, the global IPv6 adoption rate is around 36\% as of April 2022~\cite{google-ipv6} and increases rapidly. In addition to its huge address space, IPv6 is extensible with the well-known Extension Header (EH) mechanism, making it possible for innovations on top of it (e.g., SRv6~\cite{rfc-srv6}). Our design takes advantage of the EH. 

An Autonomous System (AS) is a single administrative entity or domain under the control of an Internet Service Provider (ISP). Uniquely identified by an AS number, each AS has combined routing logic and common routing policies. The ASes are interconnected through peering relationship with several hierarchical levels. They use Border Gateway Protocol (BGP) as the EGP; Within each AS, an IGP (e.g., OSPF and IS-IS) is used. Jointly, a global routing table can be constructed on each router.

The AS or domain sets a clear boundary for responsibility. The service provider of a domain can make decision on peering agreement and various routing policies. Since the intra-domain network infrastructure and routing are under control of a domain, theoretically, the service providers are jointly capable of enforcing any routing/forwarding treatments and offering premier services to selected users, as long as the services can be audited and appreciated.  
For example, each domain can commit to certain Service Level Agreement (SLA), and hence all the domains on a path can jointly achieve some level of predictable QoS guarantee. If the domains are made explicit in a packet, it becomes easy to audit their SLA compliance through various OAM tools. With such a provision, both the end users and the domain service providers are motivated to participate in the end-to-end service with QoS guarantee, enabling performance-sensitive and demanding applications over the Internet. 

As shown in Figure~\ref{fig:stack}, we make domain an explicit L3.5 protocol layer in IPv6 networks. Specifically, the domain-related information will be encapsulated in an IPv6 Routing Header (RH) extension, which is located after the IPv6 base header and before the transport layer header (or other EHs if any).

\begin{figure}[tb]
\centering
\includegraphics[width=0.5\textwidth]{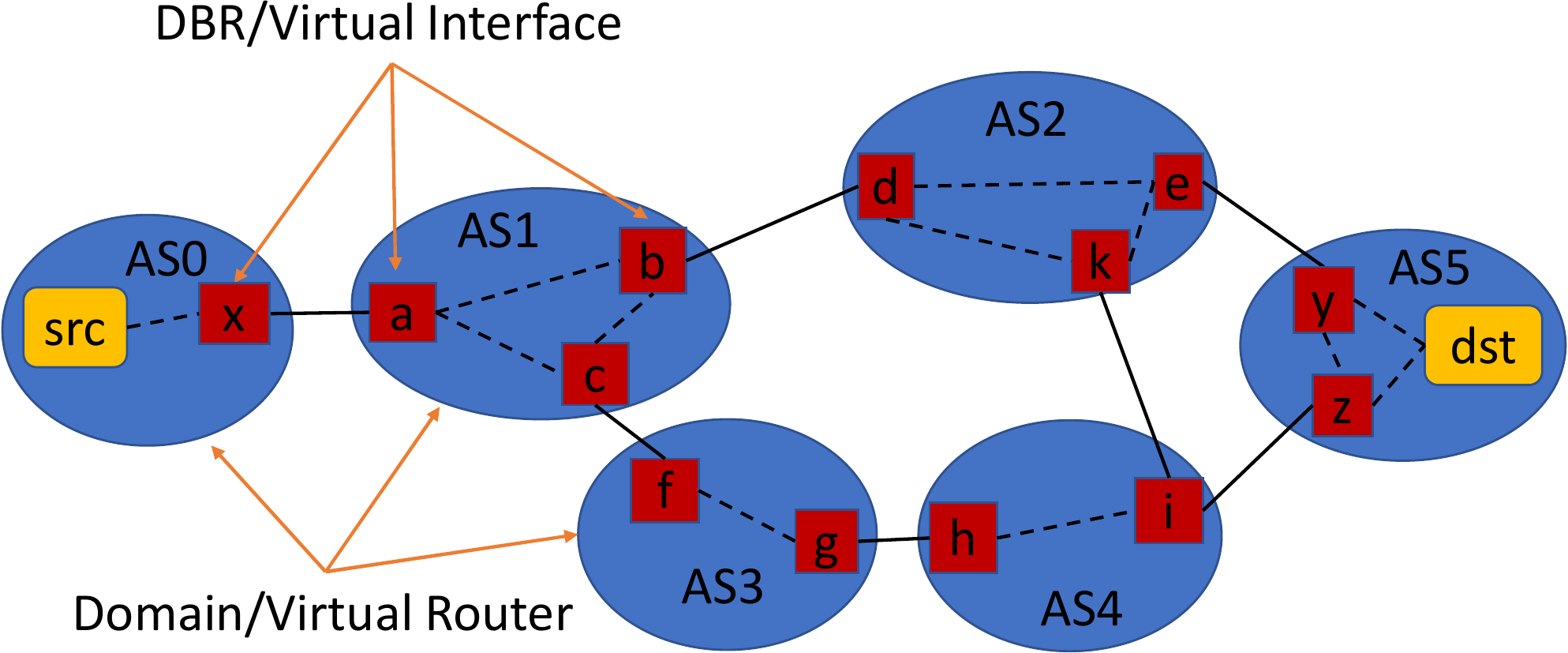}
\caption{Domain Level Routing (DLR) network architecture abstraction.}
\label{fig:network}
\end{figure}

While the whole Internet architecture remains the same, we need a new model, as illustrated in Figure~\ref{fig:network}, to describe it. 
In the new model, each domain is abstracted as a virtual router. The Domain Border Routers (DBR) are abstracted as the virtual router interface, and all the intra-domain routers of a domain form the switch fabric of the corresponding virtual router. A packet from its source to destination will be forwarded though a sequence of virtual routers. There can be multiple paths available. Each virtual router's performance is individually monitored and measured. Hence each path's performance can be inferred. Here we are particularly interested in the case of source routing, in which each packet can designate the sequence of virtual routers (i.e., domains) as the forwarding path. Presumably, such arrangement would only apply to some high-value and performance-sensitive applications, for which the users would like to pay a premium. Each virtual router (i.e., a domain's provider) needs to commit to certain SLA to be qualified as a participant, and if selected, share the profit after offering the service.    

In this model, each virtual router is identified by the domain ID (i.e., the AS number) and reachable by the IP addresses of its ingress interfaces (i.e., DBRs). Only the ingress and egress DBRs will be aware of and process the L3.5 header, simulating the packet processing engine on router line cards. To ensure the committed QoS, each virtual router is responsible for its own traffic management. That is, a domain can adopt various means such as traffic engineering or even dedicated private network. However, such means are invisible to users and other domains, allowing each service provider to have proprietary designs and architectures for better competitive edge and profit margin.

\section{System Design}

This section provides the details to make the IPv6-based Domain Level Routing (DLR) work. We discuss two flavors of DLR-based forwarding: the first is for source routing at the domain level, named \emph{Domain Level Source Routing} (DLSR); the second is for \emph{Domain-by-Domain} (DBD) forwarding, which is similar to the conventional Hop-by-Hop (HBH) forwarding, except that the ``next hop" router is replaced with the ``next domain". 

DLSR is preferred in a centralized service model. A central service broker is responsible to contract with users and service providers, and provision and monitor the services. DBD can be used in a distributed peer-to-peer model. In this model, the QoS policies and capabilities can be communicated between peering domains. Although it is a little harder to establish end-to-end QoS guarantee in DBD, the routing and forwarding are still at the domain level, so it still easier to apportion the responsibilities and audit the compliance of each service provider.    

\subsection{Packet Format}


\begin{figure}[tb]
\centering
\includegraphics[width=0.33\textwidth]{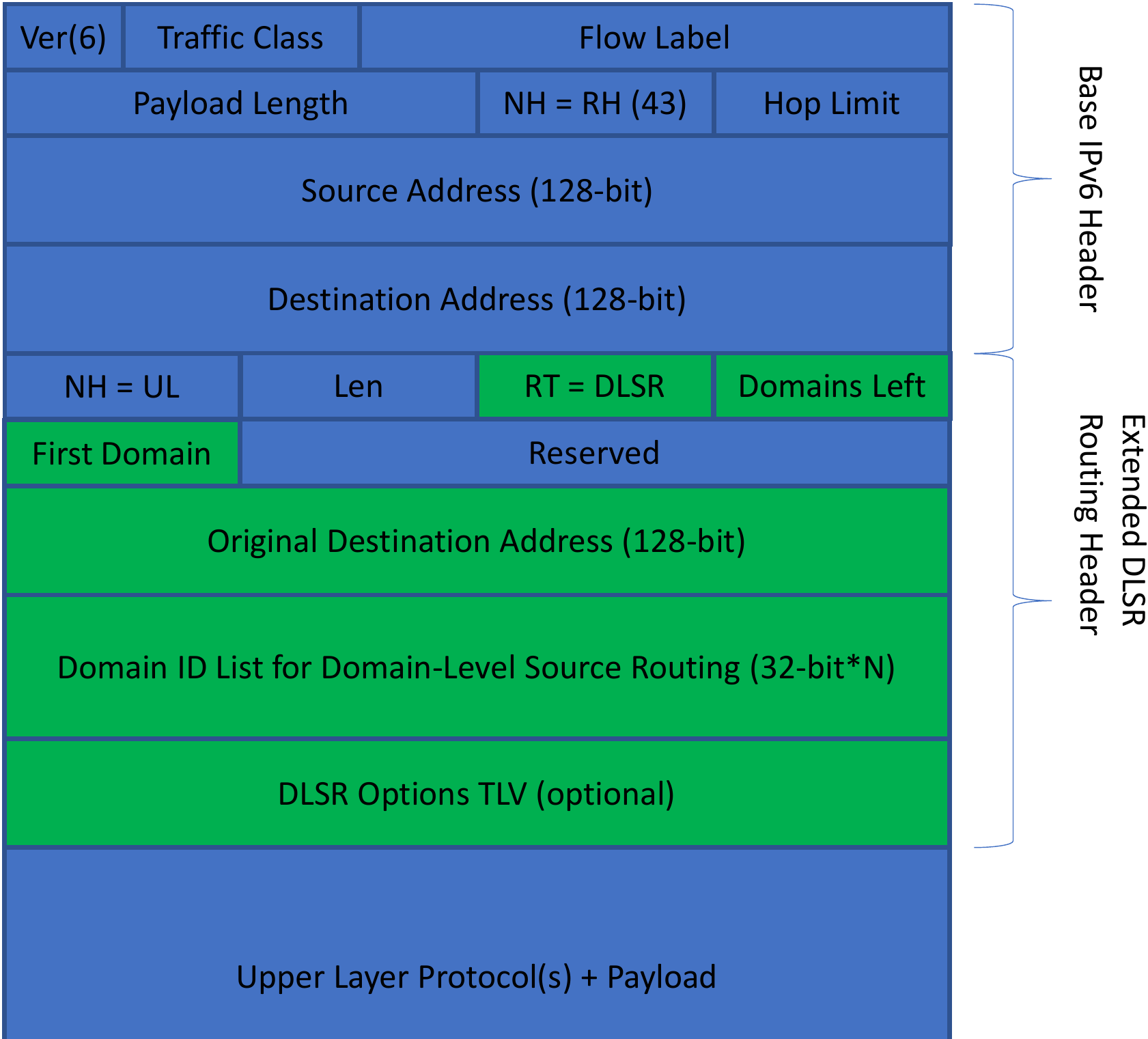}
\caption{IPv6 packet format for Domain Level Source Routing (DLSR).}
\label{fig:dlsr}
\end{figure}

\noindent\textbf{DLSR.} The packet format for DLSR is shown in Figure~\ref{fig:dlsr}. In the standard IPv6 base header, the Next Header (NH) field is set to the value 43, indicating the presence of a Routing Header (RH). In the RH extension header, the first two fields are standard EH container fields: the 8-bit NH indicates the original Upper Layer (UL) protocol type and the 8-bit \emph{Len} records the length of the EH (in the unit of 8 octets and excluding the first 8 octets). the 8-bit \emph{Routing Type} field is set to a new value (to be assigned by IANA~\cite{iana}), indicating the RH's type is DLSR. The next 8-bit field, \emph{Domain Left}, is an index to the next domain to be visited in the \emph{Domain ID List} field. The value of \emph{Domain Left} decrements by one at each domain ingress. The next 8-bit field, \emph{First Domain}, provides the index of the first entry in \emph{Domain ID List}. The following 24-bit field is reserved for other purposes such as flags.

The body of the DLSR RH is composed of three parts. The first part is used to record the original destination address. The reason for this is that on the forwarding path, the destination address in the IPv6 base header would be replaced by some address to reach a domain. Keeping the original address ensures the original packet can be recovered before the final packet delivery in the last domain. The second part is the reverse ordered Domain ID list, which defines the domain-level forwarding path. The third part is the options in Type-Length-Value (TLV) format. We will show some possible options that help enhance or facilitate the DLSR functionality. 

There can be other EHs after the DLSR RH. After the last EH is the original upper layer (i.e., transport layer) protocol header and the packet payload. 

\begin{figure}[tb]
\centering
\includegraphics[width=0.33\textwidth]{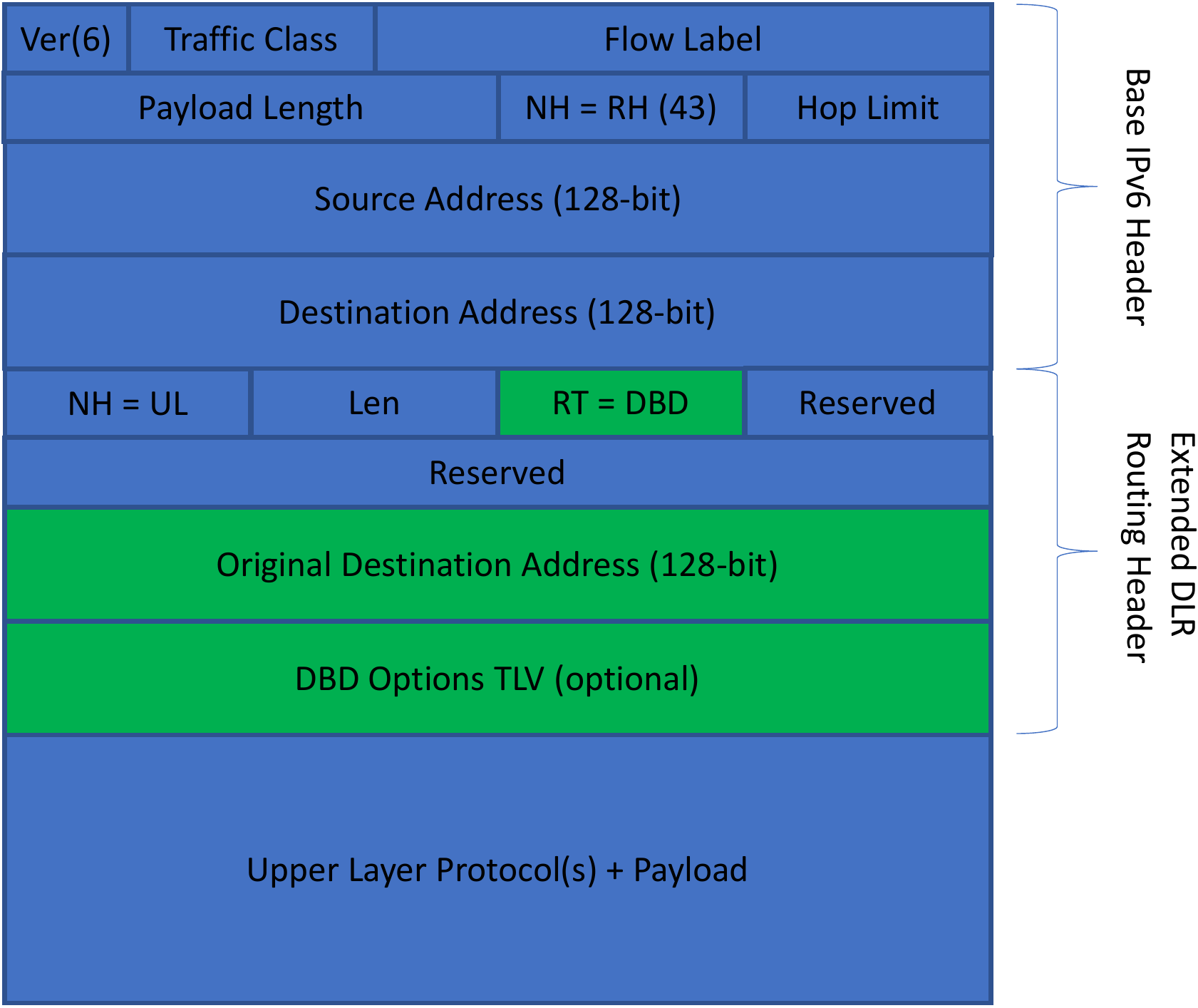}
\caption{IPv6 packet format for DLR Domain-by-Domain (DBD) forwarding.}
\label{fig:dlrh}
\end{figure}

\vspace{2mm}
\noindent\textbf{DBD.}
The DBD RH, as shown in Figure~\ref{fig:dlrh}, is simpler. It would require a new DBD routing type as well. After the first 8 octets of the DBD RH are the original destination address and DBD options in TLV format. In the DBD RH, 40 bits are reserved for future enhancement.   

\subsection{Forwarding}

\noindent\textbf{DLSR.}
Only can the ingress and egress DBRs of a domain process the RH. The other interior routers simply forward packets based on the destination address in the base header. Each DBR of domain $d$ maintains a table named Domain Entry Table (DET) that maps each peering domain $p$ of $d$ to $p$'s ingress DBR's IP address. We use Figure~\ref{fig:network} as an example to describe the forwarding process. 

Assume the source node \emph{src} in AS0 would send a packet to the destination node \emph{dst} in AS5  and it acquires a domain-level source-routing path AS0$\rightarrow$AS1$\rightarrow$AS2$\rightarrow$AS5. After the RH encapsulation, the packet's destination address is copied to the original destination address field in the RH and the domain ID list is initialized accordingly. The packet is then forwarded to the egress DBR \emph{x} of AS0. Using the \emph{First Domain} field and the Domain ID list, \emph{x} finds that it matches the first domain, AS0, on the path. \emph{x} will decrement \emph{Domains Left} and find the next domain on the path, which is AS1. AS1 is used to search the DET. DBR \emph{a}'s IP address is returned and copied into the destination address field of the base header. The packet is then forwarded to \emph{a}.

By examining the RH, the ingress DBR \emph{a} knows that AS1 is not the last domain and the next domain on the path is AS2, so it will search the DET and find \emph{d}'s address. The address is copied to the destination address field, the RH is updated (e.g., decrement \emph{Domains Left} by 1), and the packet is eventually forwarded to \emph{b}, the egress DBR of AS1. If no option needs to be processed, \emph{b} will simply forward the packet to \emph{d} according to the destination address. 

The packet is forwarded in AS2 following the similar process as in AS1 and reaches the ingress DBR \emph{y} in the last domain AS5. Once \emph{y} figures out its the ingress of the last domain, it will copy the original destination address in RH back to the destination address field in the base header. Finally, the packet is forwarded to \emph{dst} in AS5.

\vspace{2mm}
\noindent\textbf{DBD.}
In this mode, each DBR needs to maintain another table named Next Domain Table (NDT) which maps each destination network (i.e., address prefix) to a Next Domain which is used to reach the destination. NDT is actually an augmentation to the Forwarding Information Base (FIB): in addition to the Next Hop (NH), it also provides the Next Domain (ND) for a destination network. So the two tables can be merged as one to support packet forwarding with or without DBD RH.

If a packet from \emph{src} to \emph{dst} bears a DBD RH, it is forwarded to \emph{x} using the normal destination address-based forwarding process. If \emph{x} finds the packet's destination address is the same as the original destination address in the RH, it knows the packet is destined to another domain. It will search NDT and find AS1 as the next domain. Then AS1 is used to search DET and \emph{a}'s address is returned and copied to the destination address field. The packet is then forwarded to \emph{a}. 

For each ingress DBR in the following domains on the forwarding path, if it receives a DBD packet matching its address, it will check the original destination address field in the RH. If the address is in the current domain $d$ (i.e., the ND of the destination is $d$), it will be copied back to the destination address field in the based header; otherwise, the ND is acquired from NDT using the original destination address, and the new destination address is acquired from DET using the ND. The egress DBRs in the domains en route only process the options if necessary. Repeating this process, the packet is eventually delivered to \emph{dst}.  

\subsection{Routing}

DLR does not require any new control plane protocol. It can reuse the existing IGP protocols without any change and adapt BGP with minor extensions.

DBRs run BGP to generate the NDT and DET. The AS path is a mandatory attribute of BGP. When a DBR sends a BGP update to a peering DBR in another AS, it adds its own AS number to augment the AS path to a network represented as a prefix, so the AS path contains all the ASes that need to be traversed to reach the destination network. A standard routing information would include the Next Hop (NH), which is the interface address of the DBR of the next domain, along with a domain list, in which the first domain is the next domain. Hence, the routing information is enough to map a domain to one of its DBR's interface addresses. In our abstract network model, this is equivalent to the mapping between a virtual router and its ingress interface address. Hence, BGP can provide entire domain-level path information to help construct the domain-level network abstraction. 

To support DLSR, a centralized routing architecture using Path Computation Element (PCE)~\cite{rfc-pce} is also possible. We leave the detailed design as future work. 

\subsection{DNS}

In the case of DLSR, we need to provide some mechanism to acquire the domain-level source routing path. It is possible that the service broker can help provision the domain level paths and configure the DBR in the first domain to set up the path for specific applications. 

To achieve the true end-to-end DLSR (i.e., set up a domain-level path from \emph{src} to \emph{dst}), we can rely on DNS~\cite{rfc-dns} and adopt the Network-Aware Domain Name System (NA-DNS) architecture~\cite{id-nadns}. DNS is a distributed and hierarchical database used to acquire computer reachability means through the Internet. The data are formatted into various Resource Records (RRs). A typical use of DNS is for a host to get the IP addresses of a domain name by running the DNS protocol. We specify a new RR type for domain-level path. The service provider at the source domain can register the domain-level path RR for a target application in DNS servers. The source host can then make a DNS query for the domain-level path RR. The returned data are used to construct the IPv6 base header and the DLSR RH. 

For DBD, NA-DNS can also be used for hosts to express QoS requirement and get the corresponding destination address and RH option~\cite{id-nadns}.

\subsection{OAM}
Network Operations, Administration, and Maintenance (OAM) refers to the toolset for network performance measurement and troubleshooting. Since the end-to-end QoS requires independent domains to cooperate, DLR OAM is important to enforce the service contract. Through metering and accounting, measures can be taken to hold the domains who miss their SLA responsible and meanwhile compensate the behaving domains. OAM can also be used to find candidate domain-level paths.    

In-situ OAM (IOAM)~\cite{rfc-ioam} is an IETF standard for In-band Network Telemetry (INT). Working in several different modes, IOAM allows user packets to collect data from routers on the entire forwarding path. The data can be used to measure performance (e.g., bandwidth, delay, and delay jitter) and pinpoint issues (e.g., packet drop) on a per router basis. In our abstract network model, each domain is abstracted as a virtual router, so IOAM can be applied to collect data at the domain level. We encapsulate the IOAM header as an DLR RH option. Only the ingress and egress DBRs of a domain would process the IOAM in the DLR RH, which emulates the IOAM's behavior on a router's line cards. The telemetry data can be mutually verified to avoid possible frauds. For example, a domain's egress timestamp can be verified by the next domain's ingress timestamp through cross examination.           

It is conceivable that the other popular active and passive OAM methods (e.g., ping, traceroute, BFD, and TWAMP)~\cite{rfc-oam} can also be adapted to the abstract network model in which a whole domain becomes a single entity under scrutiny.  

\section{Use Cases}
We discuss several DLR use cases to demonstrate its potential for end-to-end service and QoS.

\vspace{2mm}
\noindent\textbf{Virtual Private Path.} For some long-lasting high-value applications, cross-domain virtual private path can be provisioned. Only the domain-level path is exposed to end users. Each participating domain can use its own means to achieve the QoS goal assigned to it, and are compensated based on its contribution. Following the abstract network model, the domain-level virtual private network is also possible.

\vspace{2mm}
\noindent\textbf{Deadline Based Forwarding.} Some applications require a tight and bounded delay. Any packet that misses the deadline becomes useless and therefore it is better to be discarded as soon as possible. To support such an application, each domain participating the forwarding can commit to certain delay target to make sure the overall delay is within the bound. During the packet forwarding, the used time by each domain and the remaining delay budget can be kept as a TLV option in the DLR RH. The missed target in one domain can be made up by some subsequent domains, as long as the helpers can be properly compensated. Once a domain decides that it is impossible for a packet to reach its destination on time, it can drop the packet to save bandwidth. With domain as the basic entity, the responsibility is clear and accountable.  

\vspace{2mm}
\noindent\textbf{Service Function Chaining (SFC).}
In addition to QoS, DLR is also good for service providers to collaborate on providing distributed network services. SFC is a general technique to initiate  an ordered set of service functions (e.g., firewall, load balancing, WAN optimization, NAT, and encryption) in network and steer the user traffic through it. Traditionally it can only be used by a single service provider. With DLR, it is easier to negotiate a plan to allocate different network functions in different domains. The service providers are compensated by the nature of the functions and the contribution of each participating domain. 
The SFC descriptor (e.g., NSH~\cite{rfc-nsh}) can be carried as a DLR RH option.  

\section{Related Work}

Many intra-domain QoS methods and standards (e.g., DiffServ~\cite{rfc-diffserv}, IntServ~\cite{rfc-intserv}, and MPLS-TE~\cite{rfc-rsvpte}) have been defined and applied. In our framework, these standard methods and other proprietary methods can be used within each domain to meet the QoS targets assigned to it.  

Several attempts were made to build some framework supporting end-to-end QoS (e.g., allow domains to exchange policies for inter-domain QoS provisioning and accounting~\cite{ebata1999inter}, setup inter-domain traffic engineering label switched paths with guaranteed quality of service (QoS)~\cite{idte2008}, enable cross-domain QoS accounting with limited trust~\cite{estan2007afiq}, and so on). Unfortunately, none of them succeeds, partially due to the lack of domain level visibility and accountability.     

We specify the DLSR RH format with reference to SRv6 SRH~\cite{rfc-srv6}. After all, they are both source routing schemes. The key difference is that SRv6 is based on the concept of ``segment", a forwarding instruction guiding packets to traverse a section of the network topology in an SR domain. Among many different segment types, the two most used are adjacency and prefix segments, which represent a single-hop tunnel and a multihop-tunnel, respectively. While SRv6 can simplify traffic engineering and network management across network domains, the entity of domain is still behind the scene, making the concerted end-to-end QoS difficult. 

Our design shares the similar spirit as NIRA~\cite{yang2007iton} by providing provider-level routes. However, we do not require the strict provider-rooted hierarchical addressing proposed in NIRA and we provide a practical IPv6 EH-based design.  

\section{Analysis and Discussion}

DLR is simple and compatible with the current Internet architecture. It is better to be deployed to service some specific high-value and performance-demanding applications as a starter. It does not significantly complicate the data plane and increase the forwarding table size: only DBRs need to do some extra packet processing and table lookup; NDT can be merged into FIB and DET is small. The control plane can rely on the existing protocols with minor extensions.   

The RH overhead of DLSR is $(24+4n)$ bytes where $n$ is the number of domains on a path. A 10-domain path incurs only a 64-byte RH overhead. As a comparison, a 10-segment SRv6 path would incur a 168-byte RH overhead. 

Our initial analysis shows that such a design is straightforward and incrementally deployable. However, many questions remain to be answered. For example, what is the best possible business model built on top of DLR? Is a centralized model or a peer-to-peer model for the service contract more proper? What options and fields should be defined to enhance the DLR functionality? We expect our preliminary work can trigger more discussions and detailed designs in the research and Internet standard communities.

In our future work, in order to promote standard adoption, we will prototype the system using P4~\cite{bosshart2014p4}, evaluate the performance impact of the RH processing, and demonstrate the use cases on networks built with programmable devices. 

\section{Conclusion}

The requirement for end-to-end quality of service is becoming increasingly pressing, driven by many high-value and performance-demanding applications. To keep the Internet relevant for these applications, the barriers for inter-domain QoS must be leveled. Making domain explicit in packet forwarding is an important first step towards a viable business model for end-to-end QoS and other value-added services. We show that this is achievable under the IPv6 standard framework. Without needing a heavy lifting, domain can be introduced as a new L3.5 in the protocol stack, allowing the Internet service providers to coordinate towards some common QoS and service goals with accountable responsibility.

\vspace{2mm}
\bibliographystyle{ACM-Reference-Format}
\bibliography{reference}

\end{document}